\documentclass{article}

\def\[{\left\lbrack}
\def\]{\right\rbrack}

\def\({\left(}
\def\){\right)}
\def\ih{\'\i}

\begin{document}
\date{}
\title{An Improved Gauge Unfixing Formalism and the Abelian Pure Chern Simons Theory}
\author{Jorge Ananias Neto\\
Departamento de F\ih sica, ICE, \\ Universidade Federal de Juiz de Fora, 36036-900,\\ Juiz de Fora, MG, Brazil }

\maketitle

\begin{abstract}
We propose a new scheme of embedding constrained systems based on the Gauge Unfixing formalism. Our aim is to  modify directly the original phase space variables of a system in order to be gauge invariant quantities. We apply our procedure in a nontrivial constrained model
that is the Abelian Pure Chern Simons Theory where new results are obtained. Among them we can cite  the development of a systematic procedure in order to separate the first and the second class constraints, and  the obtainment of the same initial Abelian Pure Chern Simons Lagrangian as the gauge invariant Lagrangian. This last result shows that the gauge symmetry of the action is certainly preserved.
\end{abstract}
\vskip .5 cm
\noindent PACS: 11.10.Ef; 11.15.-q\\
Keywords: constrained systems; embedding systems; gauge invariant Hamiltonians
\newpage

\section{Introduction}
The Abelian Pure Chern Simons (CS) Theory is a mixed constrained system where one of their four constraints must be redefined in order to be a first class one. Then, after this step, we have well defined algebras of two first class constraints and two second class constraints. The BFT formalism\cite{BFFT,embed}, which enlarges the phase space variables with the introduction of the Wess Zumino (WZ) fields, has been used with the objective to embed the CS  theory\cite{pcs}. As a result, the authors show many important features. Another work\cite{ncs} has also employed  the BFT formalism to study a Nonabelian version of the CS theory. In this article, the authors propose two methods that overcome the problem of embedding mixed constrained systems. In an opposite side of the BFT formalism, there is another method that embeds second class constrained systems, called Gauge Unfixing (GU) formalism. It was proposed by Mitra and Rajaraman\cite{MR} and continued by Vytheeswaran\cite{Vyt,JAN}. This formalism considers part of the total second class constraints as the gauge symmetry generators while the remaining ones form the gauge fixing terms. The second class Hamiltonian must be modified in order to satisfy a first class algebra with the constraints initially chosen to be the gauge symmetry generators. This approach has an elegant property that does not extend the phase space with extra variables. 

The purpose of this paper is to give a alternative scheme for the GU formalism and to apply this method to the CS theory. Our aim is to redefine the original phase space variables of a constrained system, without to introduce any WZ terms, in order to be gauge invariant fields. Then, after this procedure, we will construct functions of these gauge invariant fields which will be gauge invariant quantities. As we will see, we begin with a mixed constrained system that is the CS theory and, applying our formalism, we obtain a first class system written only in terms of the original phase space variables with many novel features. As many important constrained systems have only two second class constraints, then, in principle,  we present our formalism only for systems with two second class constrains without any loss of generality. In order to clarify the exposition of the subject, this paper is organized as follows: in Section 2, we give a short review of the usual GU formalism. In Section 3, we present our formalism. In Section 4, we apply our procedure to the CS theory. In Section 5, we make our concluding remarks.

\section{A brief review of the Gauge Unfixing formalism}

Let us consider a constrained system described by the second class Hamiltonian $H$ and two second class constraints $T_1$ and $T_2$. The basic idea of the GU formalism is to select one of the two second class constraints to be the gauge symmetry generator. As example, if we choose $T_1$ as the first class constraint, then, we need to scale $T_1$ as $\,\frac{T_1} {\Delta_{12}}\equiv\tilde{T}\,$
where $\Delta_{12}=\{T_1,T_2\}\,$. The second class constraint $T_2$ will be discarded. The Poisson bracket between $\tilde{T}$ and $T_2$ is $\,\{\tilde{T},T_2\}=1\,$,
so that $\tilde{T}$ and $T_2$ are canonically conjugate. The second class Hamiltonian 
must be modified in order to satisfy a first class algebra. Then, the gauge invariant Hamiltonian is constructed by the series in powers of $T_2$

\begin{equation}
\label{gh}
\tilde{H}=H+T_2\,\{H,\tilde{T}\}+\frac{1} {2!}T_2^2\, 
\{ \{H,\tilde{T}\},\tilde{T}\}+
\frac{1} {3!} T_2^3\,\{\{\{ H,\tilde{T}\},\tilde{T}\},\tilde{T}\}+\ldots,
\end{equation}
where we can show that $\{\tilde{H},\tilde{T}\}=0$ and $\tilde{T}$ must satisfy a first class algebra $\{\tilde{T},\tilde{T}\}=0$. The gauge invariant Hamiltonian, Eq.(\ref{gh}),
can be elegantly written in terms of a projection operator on the second class Hamiltonian $H$

\begin{equation}
\tilde{H}= e^{T_2 \tilde{T}_{op}}:H,
\end{equation}
where $\tilde{T}_{op}\, H\equiv\{H,\tilde{T}\}$ and an ordering prescription must be adopted that is $T_2$ must come before the Poisson bracket.

\section{The improved Gauge Unfixing formalism}
Let us start with the original phase space variables written as

\begin{equation}
F=(q_i,p_i),
\end{equation}
where $\, F \,$ can describe a particle or field model. As we haven seen in Section 2, the usual GU formalism embeds directly the second class Hamiltonian. Thus, our strategy is to construct a gauge invariant function  $\tilde{A}\,$  from the second class function  $A\,$  by gauging the original phase space variables, using for this the idea of the GU formalism. 

Denoting the first class variables by

\begin{equation}
\tilde{F}=(\tilde{q_i},\tilde{p_i}),
\end{equation}
we determine the first class function $\tilde{F}$ in terms of the original phase space variables by employing the variational condition 

\begin{equation}
\label{vc}
\delta\tilde{F}=\epsilon \{\tilde{F},\tilde{T}\}=0,
\end{equation}
where $\tilde{T}$ is the scaled second class constraint chosen to be the gauge symmetry generator and $\epsilon$ is an infinitesimal parameter. Any function of $\tilde{F}\,$ will be gauge invariant since

\begin{equation}
\{\tilde{A}(\tilde{F}),\tilde{T}\}=\{\tilde{F},\tilde{T}\} \frac {\partial\tilde{A}} {\partial\tilde{F}}=0,
\end{equation}
where

\begin{equation}
\{\tilde{F},\tilde{T}\} \frac {\partial\tilde{A}} {\partial\tilde{F}}\equiv\{\tilde{q}_i,\tilde{T}\}\frac {\partial\tilde{A}}{\partial\tilde{q}_i}+ \{\tilde{p}_i,\tilde{T}\}\frac {\partial\tilde{A}}{\partial\tilde{p}_i}.
\end{equation}
Consequently, we can obtain a gauge invariant function from the replacement of

\begin{equation}
\label{sub}
A(F)\Rightarrow A(\tilde{F})=\tilde{A}(\tilde{F}).
\end{equation}
The gauge invariant phase space variables $\tilde{F}$ are constructed by the series in powers of $T_2$

\begin{equation}
\label{ff}
\tilde{F}=F+ \sum_{n=1}^{\infty} c_n\,T_2^n=F+c_1\,T_2+c_2\,T_2^2+\ldots,
\end{equation}
where this series has an important boundary condition that is

\begin{equation}
\label{bon}
\tilde{F}(T_2=0)=F.
\end{equation}
The condition above and the relation (\ref{sub}) show that when we impose the discarded  constraint $T_2$ equal to zero, we reobtain the original second class system. Therefore, the relations (\ref{sub}) and (\ref{bon})  guarantee the equivalence between our first class model and the initial second class system.

The coefficients $c_n$ in the relation (\ref{ff}) are then determined by the variational condition, Eq.(\ref{vc}). The general equation for $c_n$ is

\begin{equation}
\delta\tilde{F}=\delta F +\sum_{n=1}^\infty\, (\delta c_n\, T_2^n+ n\, c_n\,T_2^{(n-1)}\delta T_2)=0,
\end{equation} 
where

\begin{eqnarray}
\delta F &=&\epsilon \{F,\tilde{T}\},\\
\delta c_n &=&\epsilon \{c_n,\tilde{T}\},\\
\label{res}
\delta T_2&=&\epsilon \{T_2,\tilde{T}\} = -\,\epsilon.
\end{eqnarray}
In Eq.(\ref{res}) we assume that $\{\tilde{T},T_2\}=1$. Then, for the linear correction term $(n=1)$, we have

\begin{equation}
\label{c_1}
\delta F+ c_1\,\delta T_2=0 \;\Rightarrow \;\delta F - c_1\,\epsilon=0 \;\Rightarrow\; c_1=\frac{\delta F}{\epsilon}.
\end{equation}
For the quadratic correction term (n=2), we get

\begin{equation}
\label{c_2}
\delta c_1+2c_2\,\delta T_2=0 \;\Rightarrow\; \delta c_1-2c_2\epsilon=0 \;\Rightarrow\; c_2=\frac{1} {2}\frac{\delta c_1}{\epsilon}.
\end{equation}
For $n\geq 2$, the general relation is

\begin{eqnarray}
\label{c_n}
\delta c_n+(n+1)c_{n+1}\,\delta T_2=0\;\Rightarrow\; \delta c_n-(n+1)\,c_{(n+1)}\epsilon=0\nonumber\\ \nonumber\\ \Rightarrow\; c_{(n+1)}=\frac{1}{(n+1)}\frac{\delta c_n}{\epsilon}.
\end{eqnarray}
Using the relations (\ref{c_1}), (\ref{c_2}) and (\ref{c_n}) in Eq.(\ref{ff}) we obtain the series which determines $\tilde{F}$

\begin{equation}
\label{series}
\tilde{F}=F+ T_2\,\frac{\delta F}{\epsilon}+\frac{1}{2!}T_2^2\,\frac{\delta\delta F}{\epsilon^2}+\frac{1}{3!}T_2^3\,\frac{\delta\delta\delta F}{\epsilon^3}+\ldots\,.
\end{equation}
The expression $\tilde{F}$ can also be elegantly written in terms of a projection operator on $F$

\begin{equation}
\tilde{F}=e^{T_2\,\frac{\delta}{\epsilon}}:F,
\end{equation}
where again an ordering prescription must be adopted that is $T_2$ must come before $\frac{\delta}{\epsilon}$. Now, if we calculate the Poisson bracket between the two gauge invariant variables defined by the formula (\ref{series}) and next we take the limit $T_2 \rightarrow 0$,  we get

\begin{eqnarray}
\{\tilde{F},\tilde{G}\}_{T_2=0}&=&\{F,G\}+\{F,\tilde{T}\}\{T_2,G\}-\{F,T_2\}\{\tilde{T},G\}\nonumber\\
&=&\{F,G\}+\{F,T_i\}\epsilon^{ij}\{T_j,G\}.
\end{eqnarray}
If we assume that $\{T_i,T_j\}\equiv \Delta_{ij}=\epsilon_{ij}$, being $T_1\equiv\tilde{T}$, we can write

\begin{eqnarray}
\label{dirac}
\{\tilde{F},\tilde{G}\}_{T_2=0}&=&\{F,G\}+\{F,T_i\}\Delta^{ij}\{T_j,G\}\nonumber\\
&=&\{F,G\}_D,
\end{eqnarray}
where $\Delta^{ij}\equiv\epsilon^{ij}$ is the inverse of $\Delta_{ij}\equiv\epsilon_{ij}$ and $\{F,G\}_D$ is the Dirac bracket\cite{Dirac}. Thus, we can observe that when we change the gauge invariant variables by imposing the condition $T_2=0$, we return to the original second class system where the Poisson brackets transform in the Dirac brackets. This important result also confirms the consistency of our formalism. The same result was obtained by employing the BFT formalism\cite{pcs}.

\section{The Abelian Pure Chern Simons Theory}
The CS theory, being a (2+1) dimensional field theory, is governed by the Lagrangian 

\begin{equation}
\label{initial}
L = \int d^2x\,\frac {k} {2}  \,\epsilon_{\mu\nu\rho}\, A^\mu\partial^\nu A^\rho,
\end{equation}
where $\,k\,$ is a constant. From the standard Dirac constrained formalism\cite{Dirac} we obtain three canonical momenta which are the primary constraints
\begin{eqnarray}
\label{pi0}
T_0\equiv\pi_0\approx 0,\\
\label{pii}
T_i\equiv\pi_i-\frac{k}{2}\, \epsilon_{ij} A^j\approx 0 \;\;(i=1,2).
\end{eqnarray}
Using the Legendre transformation we derive the canonical Hamiltonian 
\begin{equation}
\label{hc}
H_c= -k \int d^2 x \,A^0\,\epsilon_{ij}\,\partial^iA^j.
\end{equation}
From the temporal stability condition of the constraint, Eq.(\ref{pi0}), we get the secondary constraint
\begin{equation}
\label{pi3}
T_3\equiv k \, \epsilon_{ij}\,\partial^i A^j\approx 0.
\end{equation}
We observe that no further constraints are generated via this iterative procedure.  $T_0,T_i$ and $T_3$ are the total constraints of the model .
In order to separate the second and the first class constraints, we need to redefine the constraint (\ref{pi3}). In principle, we can suggest an expression for the constraint as (an educated guess) 

\begin{eqnarray}
\label{bomega3}
\tilde{T}_3\equiv T_3+\partial^i T_i
=\partial^i\pi_i+\frac{k}{2}\,\epsilon_{ij}\,\partial^iA^j.
\end{eqnarray}
Then, $T_0$ and $\tilde{T}_3$ form the first class constraints, while
 $T_i$, Eq.(\ref{pii}), forms the second class constraints satisfying the algebra

\begin{equation}
\{T_i(x),T_j(y)\}=-k\,\epsilon_{ij}\,\delta^3(x-y) \;\; (i,j=1,2).
\end{equation}
\vskip .5cm
Our formalism begins by choosing the symmetry gauge generator as

\begin{equation}
\label{g1}
\tilde{T}=-\frac{T_1}{k}=-\frac{\pi_1}{k}+\frac{A^2}{2}.
\end{equation}
We would like to note that the superscript in the Eq.(\ref{g1}) indicates a vector component and not an exponent. Then, we have the algebra $\{\tilde{T}(x),T_2(y)\}=\delta^3(x-y)$. The second class constraint $T_2=\pi_2+\frac{k}{2} A^1$ will be discarded. The infinitesimal gauge transformations generated by symmetry generator $\tilde{T}$ are

\begin{eqnarray}
\label{da}
\delta A^i=\epsilon \{A^i(x),\tilde{T}(y)\}=
-\frac{\epsilon}{k}\,\delta^i_1\,\delta^3(x-y),\\
\label{dpi}
\delta \pi_i=\epsilon \{\pi_i(x),\tilde{T}(y)\}=
-\frac{\epsilon}{2}\,\delta_i^2\,\delta^3(x-y),\\
\delta T_2=\epsilon\, \{T_2(x),\tilde{T}(y)\}=-\epsilon \,\delta^3(x-y).
\end{eqnarray}
The gauge invariant field $\tilde{A}^i$ is constructed by the series 
in powers of $T_2$ 

\begin{equation}
\tilde{A}^i=A^i+ b_1\,T_2+b_2\,T_2^2+\ldots+b_n\,T_2^n.
\end{equation}
From the invariance condition $\delta\tilde{A}^i=0$, we can compute all the correction terms $b_n$. For the linear correction term in order of $T_2$, we get

\begin{eqnarray}
\delta A^i+b_1\delta T_2=0 \;\Rightarrow\; -\frac{\epsilon}{k}\,\delta^i_1\delta^3(x-y)-b_1\epsilon\,\delta^3(x-y)=0 \;\Rightarrow\; b_1=-\frac{1}{k}\,\delta^i_1.
\end{eqnarray}
For the quadratic term, we obtain $b_2=0$, since $\delta b_1=\epsilon\{b_1,\tilde{T}\}=0.$ Due to this, all the correction terms $b_n$ with $n\geq 2$ are null. Therefore, the gauge invariant field $\tilde{A}^\mu$ is

\begin{eqnarray}
\label{a0}
\tilde{A}^0&=&A^0,\\
\label{ai}
\tilde{A}^i&=&A^i-\frac{1}{k}\,\delta^i_1\,T_2,
\end{eqnarray}
or
\begin{eqnarray}
\label{A0}
\tilde{A}^0&=&A^0,\\
\label{A1}
\tilde{A}^1&=&A^1-\frac{1}{k}\,T_2,\\
\label{A2}
\tilde{A}^2&=&A^2,
\end{eqnarray}
where by using Eq.(\ref{da}), it is easy to show that $\delta\tilde{A}^\mu=0$. The gauge invariant field $\tilde{\pi}_i$ is also constructed by the series in powers of $T_2$ 

\begin{equation}
\tilde{\pi}_i=\pi_i+ c_1\,T_2+c_2\,T_2^2+\ldots+c_n\,T_2^n.
\end{equation}
From the invariance condition $\delta\tilde{\pi}_i=0$, we can compute all the correction terms $c_n$. For the linear correction term in order of $T_2$, we get

\begin{eqnarray}
\delta \pi_i+c_1 \delta T_2=0 \;\Rightarrow\; -\frac{\epsilon}{2}\delta_i^2\delta^3(x-y)-c_1\epsilon\,\delta^3(x-y)=0 \,\Rightarrow\,  c_1=-\frac{1}{2}\delta_i^2.
\end{eqnarray}
For the quadratic term, we obtain $c_2=0$, since $\delta c_1=\epsilon\{c_1,\tilde{T}\}=0.$ Due to this, all the correction terms $c_n$ with $n\geq 2$ are null. Therefore, the gauge invariant field $\tilde{\pi}_i$ is

\begin{equation}
\tilde{\pi}_i=\pi_i-\frac{1}{2}\,\delta_i^2\,T_2,
\end{equation}
or
\begin{eqnarray}
\tilde{\pi}_1&=&\pi_1,\\
\tilde{\pi}_2&=&\pi_2-\frac{1}{2}\,T_2,
\end{eqnarray}
where by using Eq.(\ref{dpi}), it is easy to show that $\delta\tilde{\pi}_i=0$. The Poisson brackets between the gauge invariant fields are

\begin{eqnarray}
\label{aiaj}
\{\tilde{A}^i(x),\tilde{A}^j(y)\}=\frac {1}{k}\,\epsilon^{ij}\,\delta^3(x-y),\\
\label{pipj}
\{\tilde{\pi}_i(x),\tilde{\pi}_j(y)\}=\frac {k}{4}\,\epsilon_{ij}\,\delta^3(x-y),\\
\label{aipij}
\{\tilde{A}^i(x),\tilde{\pi}_j(y)\}=\frac {1}{2}\,\delta^i_j\,\,\delta^3(x-y).
\end{eqnarray}
We can observe that the Poisson brackets, Eqs.(\ref{aiaj}),(\ref{pipj}) and (\ref{aipij}), reduce to the original Dirac brackets\cite{pcs} since $T_2=0$, as discussed in Eq.(\ref{dirac}). The gauge invariant Hamiltonian, written only in terms of the original phase space variables, is obtained by substituting $A^\mu$ by $\tilde{A}^\mu$, Eqs.(\ref{a0}) and (\ref{ai}), in the canonical Hamiltonian, Eq.(\ref{hc}), as follows

\begin{eqnarray}
\label{fh}
\tilde{H}=k \int d^2x\ \,\partial^i\tilde{A}^0\,\epsilon_{ij}\,\tilde{A}^j=
H_c+\int d^2x \;\partial^2A^0\;T_2\nonumber\\ \nonumber\\
=\int d^2x\;[k\, \partial^iA^0\, \epsilon_{ij}\, A^j + \partial^2 A^0\,\pi_2+\frac{k}{2}\,\partial^2 A^0\, A^1].
\end{eqnarray}
Imposing the temporal stability condition of $\pi_0\,(T_0\equiv\pi_0)$

\begin{eqnarray}
\label{hpi0}
\{\pi_0,\tilde{H}\}=0 \; \Rightarrow \; k \, \epsilon_{ij}\,\partial^i A^j+\partial^2\pi_2+\frac{k}{2}\partial^2A^1= k \, \epsilon_{ij}\,\partial^i A^j+\partial^2 T_2=0
\nonumber\\ \nonumber\\ \Rightarrow k \; \epsilon_{ij}\,\partial^i \tilde{A}^j=0,
\end{eqnarray}
we get the secondary constraint

\begin{equation}
\tilde{T}_3\equiv k \, \epsilon_{ij}\,\partial^i \tilde{A}^j,
\end{equation}
that is just the secondary constraint, Eq.(\ref{pi3}), with the replacement of $A^i$ by $\tilde{A}^i$. The gauge invariant Hamiltonian $\tilde{H}$ and the irreducible constraints $T_0,\tilde{T} $ and $\tilde{T}_3$ form a set of first class algebra given by

\begin{eqnarray}
\{\tilde{H},\tilde{T}\}=0,\\
\{\tilde{H},T_0\}=\tilde{T}_3,\\
\label{ht3}
\{\tilde{H},\tilde{T}_3\}=0,\\
\label{tt3}
\{\tilde{T},\tilde{T}_3\}=0,\\
\{\tilde{T},T_0\}=0,\\
\{T_0,\tilde{T}_3\}=0,
\end{eqnarray}
where we have used relation (\ref{aiaj}) to prove Eq.(\ref{ht3}) and the condition $\,\delta \tilde{A}^i=0\,$ to prove Eq.(\ref{tt3}). Here, we would like to mention important results obtained by our formalism. First, by imposing the temporal stability of $T_0$, Eq.(\ref{hpi0}), we get, by a systematic way, an irreducible first class constraint $\tilde{T}_3$. Second, we only embed  the initial second class constraint $T_1$, Eq.(\ref{pii}), and, consequently, we have all the constraints forming a first class set.  Moreover, in order to reduce all the constraints of the CS theory in a second class nature it is enough to assume $T_2=0$. 

Finally, the gauge invariant CS Lagrangian can be deduced by performing the inverse Legendre transformation

\begin{equation}
\label{fl}
\tilde{L}=\int d^2x\; (\,\tilde{\pi}_i\dot{\tilde{A}^i}-\tilde{H}\,),
\end{equation}
where $\tilde{H}$ is given by Eq.(\ref{fh}). As the gauge invariant Hamiltonian, $\tilde{H}$, has the same functional form of the canonical Hamiltonian, Eq.(\ref{hc}),
thus, from the inverse Legendre transformation, Eq.(\ref{fl}), we can deduce that the first class Lagrangian (written in terms of the first class variables) will take the same functional form of the original Lagrangian, Eq.(\ref{initial})

\begin{equation}
\label{fl2}
\tilde{L}=\int d^2x\,\frac {k} {2}  \,\epsilon_{\mu\nu\rho}\, \tilde{A}^\mu\partial^\nu \tilde{A}^\rho.
\end{equation}
Using the Eqs.(\ref{A0}), (\ref{A1}) and (\ref{A2}), the gauge invariant Lagrangian, Eq.(\ref{fl2}), becomes
\begin{eqnarray}
\label{fl22}
\tilde{L}=\int d^2x\,\frac {k} {2}\, [\, A^0\partial^1A^2-A^0\partial^2 A^1+\frac{1}{k}\,A^0\partial^2 T_2\nonumber\\+A^1\partial^2A^0-\frac {1}{k}T_2\,\partial^2A^0-A^1\partial^0A^2+\frac{1}{k}T_2\,\partial^0A^2\nonumber\\+A^2\partial^0A^1- \frac{1}{k}A^2\partial^0T_2-A^2\partial^1A^0\,]\,.
\end{eqnarray}
The Hamilton equation of motion produces a relation for $\partial^0 A^2$ given by

\begin{equation}
\label{a2}
\partial^0 A^2=\{A^2,\tilde{H}\}=\partial^2 A^0.
\end{equation}
Then, using the Eq.(\ref{a2}) and integrating by parts in the first class Lagrangian, Eq.(\ref{fl22}), we obtain

\begin{eqnarray}
\label{gl}
\tilde{L}=\int d^2x\,\frac {k} {2}\, [\, A^0\partial^1A^2-A^0\partial^2 A^1\nonumber\\+A^1\partial^2A^0-A^1\partial^0A^2\nonumber\\
+A^2\partial^0A^1-A^2\partial^1A^0\,]
\nonumber\\\nonumber\\
=\int d^2x\,\frac {k} {2}  \,\epsilon_{\mu\nu\rho}\, A^\mu\partial^\nu A^\rho.
\end{eqnarray}
We can observe that the gauge invariant Lagrangian, Eq.(\ref{gl}), reduces to the original Lagrangian, Eq.(\ref{initial}). 
The relation (\ref{gl}) is also an important result because without the presence of the extra terms in the gauge invariant Lagrangian, the original gauge symmetry transformation $\;A^\mu\rightarrow A^\mu+\partial^\mu \Lambda \;$ is certainly maintained. 

\section{Conclusions}
In this paper, we have improved the GU formalism by gauging the original phase space variables of a constrained system.  In the 
case of a system with two second class constraints, one of the constraints will be chosen to form the scaled gauge symmetry generator while the other will be discarded. The discarded constraint is used to construct a series for the gauge invariant fields. Consequently, any functions of the gauge invariant fields are gauge invariant quantities. We apply our formalism to the CS model where new results are obtained.
Our improved GU formalism can also be used to study the Nonabelian version of the Chern Simons theory\cite{ncs,kp}.

\section{Acknowledgments}
This work is supported in part by FAPEMIG, Brazilian Research Agency.


\begin{thebibliography} {99}

\bibitem{BFFT}
L. Faddeev and S. L. Shatashivilli, Phys. Lett. B167, 225 (1986). I. A. Batalin and I. V. Tyutin, Int. J. Mod. Phys. A6, 3255 (1991).

\bibitem{embed}
R. Amorim and J. Barcelos-Neto, Phys. Rev. D53, 7129 (1996); 
W. Oliveira and J. Ananias Neto, Nucl. Phys. B533, 611 (1998); M. I. Park and Y. J. Park, Int. J. Mod. Phys. A13, 2179 (1998); S. T. Hong, Y. W. Kim and Y. J. Park, Phys. Rev. D59, 114026 (1999);  C. Neves and C.  Wotzasek, Phys. Rev.D59, 125018 (1999).  J. Ananias Neto, C. Neves and W. Oliveira, Phys. Rev.D63, 085018 (2001).
M. Monemzadeh and A. Shirzad, Int. J. Mod. Phys. A18, 5613 (2003), Phys. Lett. B584, 220 (2004).

\bibitem{pcs}
M. I. Park, Y. J. Park, J.Korean Phys.Soc.31, 802 (1997).

\bibitem{ncs}
M. Monemzadeh and A. Shirzad, Phys. Rev. D72, 045004 (2005).

\bibitem{MR}
P. Mitra and R. Rajaraman, Ann. Phys.(N.Y.) 203, 157 (1990). K. Harada and H. Mukaida, Z. Phys. C  Part. Fields 48, 151 (1990).

\bibitem{Vyt}
A. S. Vytheeswaran, Ann. Phys. (N.Y.) 206, 297 (1994).

\bibitem{JAN}
J. Ananias Neto, Braz. Jour. Phys. 36,1B, 237 (2006).

\bibitem{Dirac}
P. A. M. Dirac, ``Lectures on Quantum Mechanics", Dover Publications, Mineola, N.Y. (2001).

\bibitem{kp}
W. T. Kim and Y. J. Park, Phys. Lett. B336, 376 (1994).


\end{thebibliography}
\end{document}